\documentclass[%
 reprint,
 amsmath,amssymb,
 aps,
]{revtex4-2}

\usepackage{subfigure}
\usepackage{float}
\usepackage{slashed}
\usepackage{xcolor}
\usepackage{amsmath}
\usepackage{subcaption}
\usepackage[english]{babel}
\usepackage{lineno, blindtext}
\usepackage{tikz}
\usepackage{adjustbox}
\usepackage{rotating}
\usepackage{rotfloat}
\usepackage[utf8]{inputenc}
\usepackage{graphicx}
\usepackage{dcolumn}
\usepackage{bm}

\usepackage{hyperref}
\usepackage{xcolor}
\hypersetup{
  colorlinks   = true, 
  urlcolor     = red, 
  linkcolor    = blue, 
  citecolor   = blue 
}

\usepackage{threeparttable}
\begin{document}
\preprint{APS/123-QED}

\title{Probing a novel neutral heavy gauge boson within the mono-Z$^{\prime}$ portal at the HL-LHC.}
 \author{Ali Muhammad H. H.}
 \altaffiliation[Ali.Hamed@bue.edu.eg]{}
 \affiliation{Physics Department, Faculty of Science, Ain Shams University, Elsarayat St., Abbaseya, 11517, Cairo, Egypt,\\Basic Science Department, Faculty of Engineering, The British University in Egypt, P.O. Box 43, El Sherouk City, Cairo 11837, Egypt.}
\author{El-sayed A. El-dahshan}
 \affiliation{Physics Department, Faculty of Science, Ain Shams University, Elsarayat St., Abbaseya, 11517, Cairo, Egypt.}
 \author{S. Elgammal}
 \affiliation{Centre for Theoretical Physics, The British University in Egypt, P.O. Box 43, El Sherouk City, Cairo 11837, Egypt.}
\begin{abstract}
{
This study examines the production of dark matter events alongside a Z$^{\prime}$ boson decaying through leptonic channels in simulated proton-proton collisions at the Large Hadron Collider. The analysis focuses on collisions at  $\sqrt{s}$  = 14 TeV under high-luminosity conditions, corresponding to an integrated luminosity of 1000 fb$^{-1}$. Using Monte Carlo simulations interpreted within the Effective Field Theory (EFT) framework, the work investigates potential signatures of new physics. In the absence of such phenomena, upper limits are placed on critical EFT parameters, including the theory's cutoff scale and the mass of the Z$^{\prime}$ boson.
}
\begin{description}
\item[Keywords]
Beyond the Standard Model, Dark Matter, The High-Luminosity Large Hadron Collider, New Neutral Heavy Gauge Boson
\end{description}
\end{abstract}
\maketitle
\section{Introduction}
\label{sec:intro}
There is a strong belief in scientific circles, based on astronomical observations, that the known baryonic matter is not the whole story and only forms about $27\%$ of the mass of the universe \cite{R2, R3, R4, R5, R6, R7, R8, R9, R10}. 
The rest is attributed to dark energy and dark matter (DM).
This mystery of DM has been an ongoing research point for astronomers and particle physicists on equal feet for decades. Solving this mystery is one of the primary goals of the renowned experiments at CERN's Large Hadron Collider (LHC), namely the Compact Muon Solenoid (CMS) and the ATLAS detectors. 

The most prosperous theory in particle physics is the standard model of particle physics (SM), in which two of the four fundamental forces, electromagnetism, and the weak nuclear force, are unified in one entity called the electroweak (EW) force \cite{R11}. 
Although this theory has achieved great success, the last of which was the discovery of the Higgs boson at the LHC in 2012 by ATLAS and CMS collaborations \cite{R2012}, it fails to answer a wide range of questions of interest, including what particles form DM \cite{R11, R12}. 

Many theorists have proposed various theories beyond the standard model (BSM) hoping to extend the success of  SM to more than discovering the Higgs particle and being obsessed with finding  DM particles \cite{R12}. 
Some of the promising BSM models have a topology, referred to as "mono-X," and make predictions about possible occurrences involving the production of DM in association with a visible particle, which acts as a candle, single photon, quark, or gauge boson (i.e. Z, W or Higgs) \cite{R101, R102, R103, R106, R107, R108}.
These occurrences are identified by visible state particles and a substantial amount of missing transverse energy ($E_{T}^{miss}$) signifying the presence of DM.

In alternative models, X might be an unseen particle, such as the heavy neutral gauge boson Z$^{\prime}$. An example of DM models is proposed in \cite{R1} with topology mono-Z$^{\prime}$, in which DM particles can be generated alongside Z$^{\prime}$ through three potential scenarios: the light vector (LV), dark Higgs (DH), and effective field theory (EFT). This model is sensitive at 14 TeV LHC as proposed in \cite{R1000}.
Both the CMS and ATLAS collaborations have previously searched for the massive extra-neutral gauge boson, Z$^{\prime}$, which is a prediction of Grand Unified Theory (GUT) and Supersymmetry (SUSY) \cite{Extra-Gauge-bosons, gaugeboson1, LR-symmetry11, Super-symmetry12}. Despite their efforts, they found no evidence for its existence using the complete RUN-II data set from the LHC up to 5.2 TeV of the dilepton invariant mass ($M_{\mu\mu}$) \cite{zprime, zprimeATLAS}.

In addition, the ATLAS collaboration has investigated the leptonic decay of Z$^{\prime}$ \cite{R53} and its hadronic decay \cite{R55} produced in association with dark matter. They thoroughly examined the LV and DH scenarios. Neither the CMS nor the ATLAS collaboration has explored the EFT interpretation of the mono-Z$^{\prime}$ portal.
In reference \cite{R53}, they used the full RUN-II data taken by the LHC at 13 TeV center-of-mass energy ($\sqrt{s}$ ) of proton-proton collisions with an integrated luminosity ($\mathcal{L}$) of 140 fb$^{-1}$. 
Their search has ruled out Z$^{\prime}$ masses ($M_{Z^{\prime}}$) ranging from 200 to 450 GeV for the heavy dark sector (HDS) of the LV scenario, taking $g_{q} = g_{l} = 0.1$ and $g_{D} = 1.0$, where $g_{q}$, $g_{l}$, and $g_{D}$ are the couplings of  Z$^{\prime}$ to SM quarks, leptons, and DM particles, respectively. 

This investigation examines the mono-$Z^{\prime}$ topology in the muonic decay channel of the neutral heavy gauge boson Z$^{\prime}$. Monte Carlo (MC) simulation replicates proton-proton collisions at $\sqrt{s}$ = 14 TeV with $\mathcal{L}$ = 1000 fb$^{-1}$ at the LHC.
This particular range of $\mathcal{L}$ at the LHC is called the High-Luminosity Large Hadron Collider (HL-LHC), the upcoming LHC upgrade expected to commence between 2028 and 2029 \cite{R14}. The EFT scenario is the focal point of this study. This paper aims to set limits on two of the EFT free parameters, the EFT cut-off scale $\Lambda$ and $M_{Z^{\prime}}$.

The paper is structured as follows. Section \ref{section:model} briefly explains the EFT scenario in the context of mono-$Z^{\prime}$ portal. Section \ref{section:CMS_HL-LHC_Project} briefly introduces the upcoming upgrade of the LHC, known as the HL-LHC, and the CMS detector. Moving on to the methodology discussion, the simulated signal samples and their SM background sources are discussed in section \ref{section:signal&background}. Following that, in section \ref{section:EventSelection}, the event selection, and the analysis approach are covered. The yielded results and conclusion of the study are presented in section \ref{section:Results} and section \ref{section:Conclusion}, respectively.

\section{The EFT Scenario in the context of the mono-$Z^{\prime}$ portal}
\label{section:model}
The new DM model proposed in \cite{R1} suggests that DM production accompanied by a resonance yielded from a new neutral heavy gauge boson, known as the Z$^{\prime}$, is possible. This model comes in three different scenarios: the LV scenario, also known as the dark fermion scenario, the DH scenario, and the EFT which is the light vector accompanied by the coupling of the inelastic effective field theory scenario.
\begin{figure}
\centering
\subfigure[The EFT diagram]{
  \includegraphics[width=80mm]{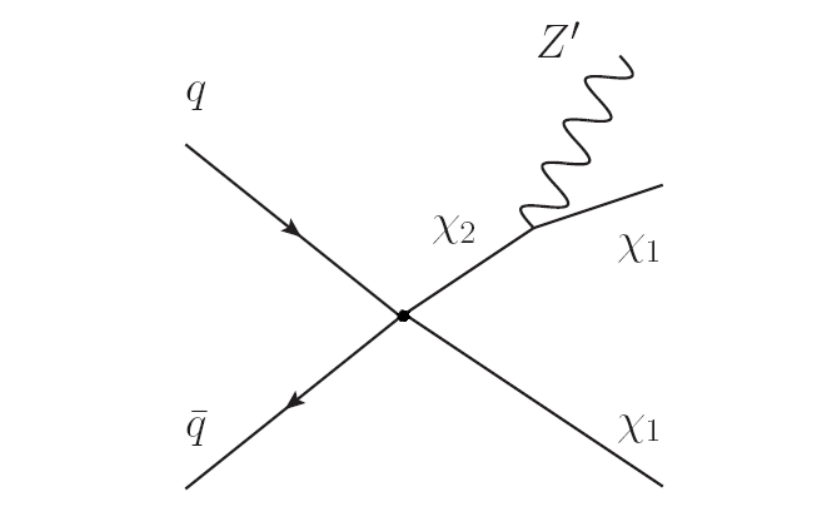}
  \label{xsec2}
}
\caption{The Feynman diagram for the EFT scenario, adapted from \cite{R1}.}
    \label{figure:fig1}%
\end{figure}
\begin{table} []    
\centering
\begin {tabular} {ll}
\hline
\hline
Scenario & \hspace{0.5cm} Masses assumptions \\
\hline
  \\ 
    Heavy dark sector & \hspace{0.5cm}  $M_{\chi_{1}} = M_{Z^{\prime}}/2$ \\
   & \hspace{0.5cm} $M_{\chi_{2}} = 2M_{Z^{\prime}}$
 \\ \\
\hline
\hline
\end{tabular}
\caption{Mass assumptions for the HDS in the EFT scenario, based on \cite{R1}.}
\label{table:tab1}
\end{table}

The EFT scenario differs from the LV scenario in that DM and SM particle interactions are reduced to a contact interaction expressed in the Lagrangian interaction term shown in equation (\ref{eq1}).
\begin{equation}
\label{eq1}
    \frac{1}{2\Lambda^{2}}\bar{q}\gamma^{\mu}q(\bar{\chi}_{2}\gamma^{\mu}\gamma^{5}\chi_{1}~+~\bar{\chi}_{1}\gamma^{\mu}\gamma^{5}\chi_{2}).
\end{equation}
Figure \ref{figure:fig1} displays the Feynman diagram for the EFT \ref{xsec2} scenario as taken from \cite{R1}, while table \ref{table:tab1} points to the dark sector mass assumption used for the process.
The primary reason for opting for the EFT scenario instead of relying on extrapolated results from the LV scenario lies in the EFT's significantly longer tail in the $E_{T}^{miss}$ distribution as explained in \cite{R1}. This distinctive feature allows it to stand out from the SM background more effectively. Additionally, this study establishes a limit on $\Lambda$, which has not been previously explored.

Throughout this paper, the coupling of SM leptons with Z$^{\prime}$ is denoted as ${\fontfamily{qcs}\selectfont{g}_{l}}$, and the coupling of DM particles with the Z$^{\prime}$ is denoted as ${\fontfamily{qcs}\selectfont{g}_{D}}$.
The main free parameters of the EFT scenario are the model $\Lambda$ and $M_{Z^{\prime}}$. Despite finding that changing ${\fontfamily{qcs}\selectfont{g}_{l}}$ changes the corresponding cross-section values, ${\fontfamily{qcs}\selectfont{g}_{l}}$ is chosen to be 0.1 to be consistent with the results of the ATLAS collaboration presented in \cite{R53}. 

The study scope is on the Z$^{\prime}$ decay to dimuon ${Z}'\rightarrow\mu\bar{\mu}$ by simulating the behavior of the CMS detector. This choice was motivated by the CMS detector's optimization for detecting the muonic decay channel of ${Z}'\rightarrow\mu\bar{\mu}$. Additionally, we examined the range of masses in the HDS, as detailed in table \ref{table:tab1}.

The EFT signal events of the model were generated using the general-purpose matrix element event generator known as \text{MadGraph5\_aMC@NLO}~v3.5.0 \cite{R33}, at 
$\sqrt{s}$ = 14 TeV.  
These generated data are used to set limits to the free parameters, $\Lambda$ and $M_{Z^{\prime}}$, as explained later in the results section.

\begin{figure*}
\centering
\subfigure[$\sigma\times Br(Z'\rightarrow \mu\mu)$ vs $M_{Z'}$]{
  \includegraphics[width=86mm]{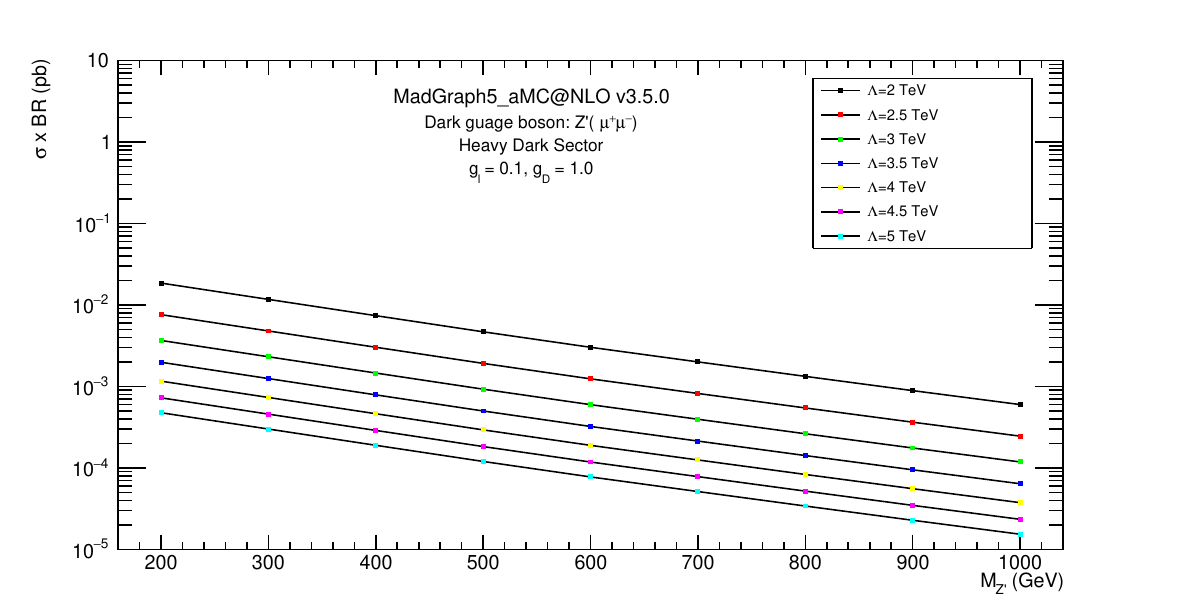}
  \label{xsec1}
}
\hspace{0mm}
\subfigure[$\sigma\times Br(Z'\rightarrow \mu\mu)$ vs $\Lambda$]{
  \includegraphics[width=86mm]{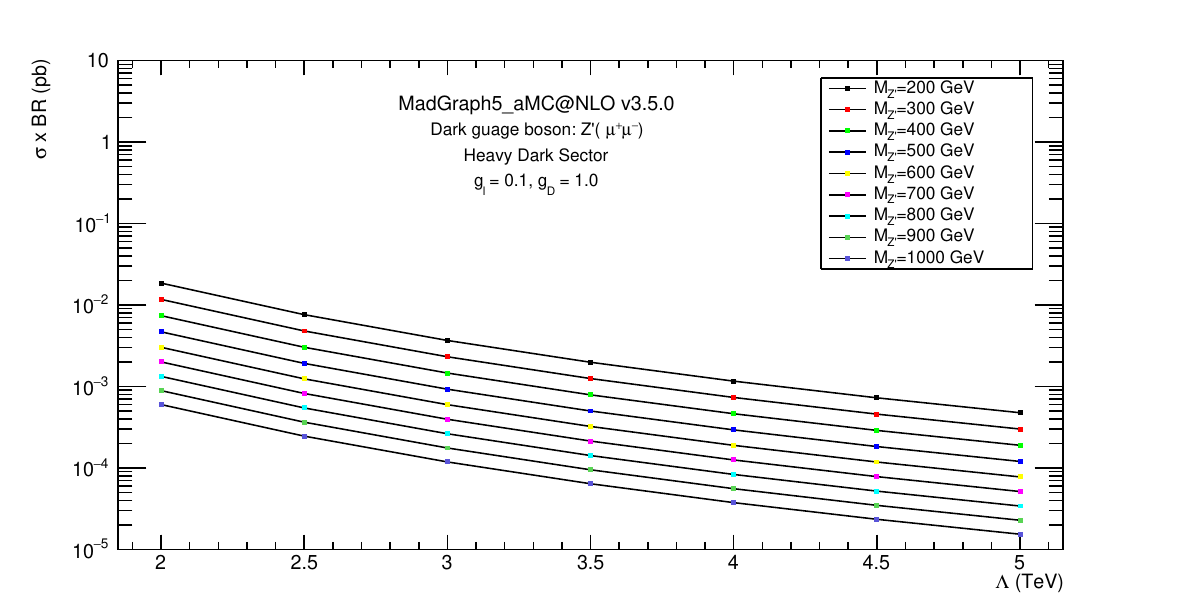}
  \label{xsec2}
}
\caption{ $\sigma \times \text{Br}(Z' \rightarrow \mu\mu)$ for EFT signal as a function of (a) $M_{Z'}$ and (b) $\Lambda$. The results are obtained for proton-proton collisions at $\sqrt{s} = 14$ TeV using \text{MadGraph5\_aMC@NLO}~v3.5.0.}
    \label{figure:fig2}%
\end{figure*}

Figure \ref{figure:fig2} presents the production of the EFT signal cross-section versus $M_{Z^{\prime}}$ \ref{xsec1} and $\Lambda$ \ref{xsec2} at $\sqrt{s}$ = 14 TeV. 
The cross-section times the branching ratio ($\sigma\times Br(Z'\rightarrow \mu\mu)$) is expected to decrease with $\Lambda$ and $M_{Z^{\prime}}$.

\section{The HL-LHC Project}
\label{section:CMS_HL-LHC_Project}
The LHC has undergone several upgrades, including Run-I, Run-II, and Run-III, making it a highly valuable scientific investment. The next phase, the HL-LHC, will further increase collision energy and frequency, enabling more precise statistical studies. This upgrade requires new equipment to be installed over 1.2 km of the 27 km long LHC \cite{R14}.

In this study, \text{Delphes} is used to model the CMS detector response. The CMS detector, a key instrument at CERN's LHC, employs a complex coordinate system for measurements, including the polar angle ($\theta$), azimuthal angle ($\phi$), and pseudo-rapidity ($\eta$). The coordinate system is defined with the $z$-axis along the beam axis, the $x$-axis toward the LHC center, and the $y$-axis upward. The pseudo-rapidity, $\eta = -\ln[\tan(\theta/2)]$, is crucial for describing particle trajectories relative to the collision point \cite{R15,R27,R28}.

The CMS detector is undergoing a significant upgrade to its silicon tracking system, replacing both the Inner Tracker (IT) and Outer Tracker (OT). The new detectors will feature enhanced radiation resistance, higher granularity, and improved data-handling capabilities. A major innovation is the introduction of a 40 MHz silicon-based track trigger in the OT, enabling real-time identification of charged particle trajectories at the hardware level, a first for CMS. This upgrade, focusing on advanced electronics for the IT and the development of the L1 track trigger, is critical for the success of the HL-LHC. Recent progress in prototype testing highlights the advancements in this area \cite{CMS:2023}.

\vspace{-16pt} 
\section{MC Samples Simulation}
\label{section:signal&background}
\subsection{Simulations of Signal Samples}
The signal events of the model were generated using \text{MadGraph5\_aMC@NLO}~v3.5.0 \cite{R33}. The next-to-leading-order (NLO) is used in cross-section calculations at $\sqrt{s} = $ 14 TeV, and \text{Pythia8} \cite{R34} is used for hadronization and parton showering processes. 
For fast detector simulation for CMS experiment, \text{Delphes} \cite{R54} has been used. The EFT production cross-section calculations have been scanned for an extensive variety of the $M_{Z^{\prime}}$ ranging from 200 GeV to 2000 GeV, and for $\Lambda$ from 1 TeV to 5 TeV assuming $g_{l} = 0.1$, and $g_{D} = 1.0$ as used in \cite{R53, R55}.
\subsection{Simulation of SM Backgrounds Samples}
In this analysis, the studied EFT signal topology is dimuon plus missing transverse energy ($\mu^{+}\mu^{-} + E^{miss}_{T}$). Therefore, several SM processes could mimic this topology via having muons and/or $E^{miss}_{T}$, arising from undetected neutrinos, in their final states. Such interactions are considered SM backgrounds for the signal. Those SM backgrounds are the Drell-Yan (DY) process ($DY\rightarrow{\mu^{+}\mu^{-}}$), the fully leptonic decay of top-quark pairs ($t\bar{t}\rightarrow{\mu^{+}\mu^{-} + 2b + 2\nu}$), the single-top process ($\bar{t}W^{+}\rightarrow{\mu^{+}\mu^{-} + 2b + 2\nu}$, $tW^{-}\rightarrow{\mu^{+}\mu^{-} + 2b + 2\nu}$), and the production of EW diboson channels ($W^{+}W^{-}\rightarrow{\mu^{+}\mu^{-} + 2\nu}$, $W^{\pm}Z\rightarrow{\mu^{\pm}\mu^{+}\mu^{-} + \nu}$, $ZZ\rightarrow{\mu^{+}\mu^{-} + 2\nu}$, and $ZZ\rightarrow{\mu^{+}\mu^{-}\mu^{+}\mu^{-}}$). All these MC Samples have a cross-section calculated at NLO using \text{MadGraph5\_aMC@NLO}~v3.5.0 \cite{R33} interfaced with \text{Pythia8} \cite{R34} for hadronization and modeling. All contributions of SM background processes and signal samples are estimated using MC Simulation and normalized to their corresponding cross-section with $\mathcal{L}$ = 1000 fb$^{-1}$. 

\section{Selection of Events}
\label{section:EventSelection}
\begin{figure*}
\centering
\subfigure[Dimuon invariant mass]{
  \includegraphics[width=86mm]{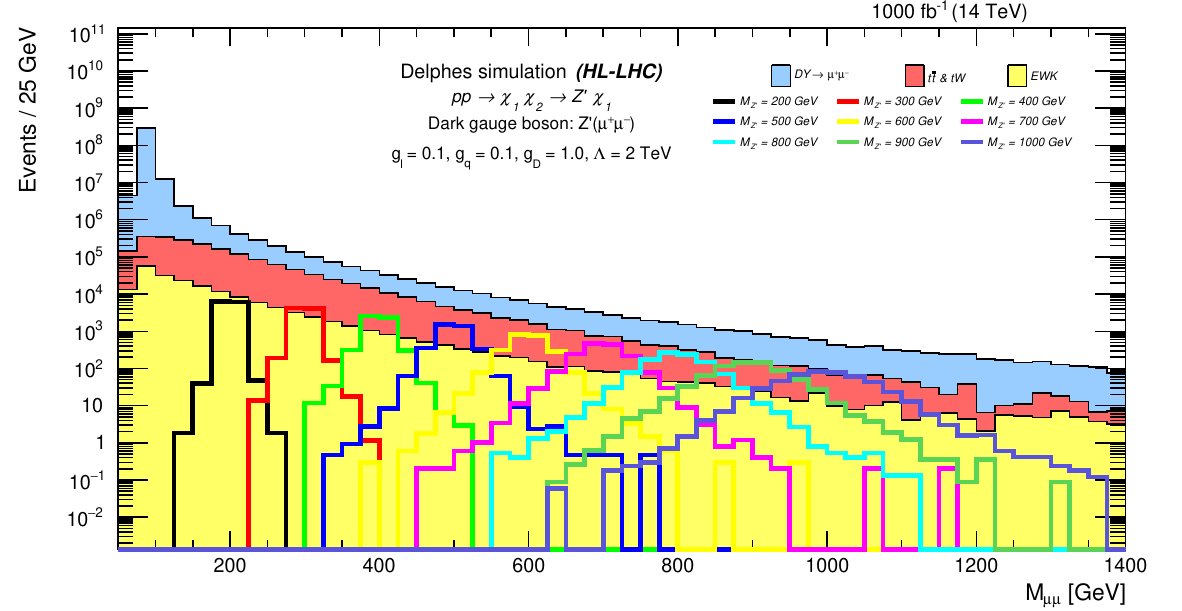}
  \label{mass}
}
\hspace{0mm}
\subfigure[Missing transverse energy ($E_{T}^{miss}$)]{
  \includegraphics[width=86mm]{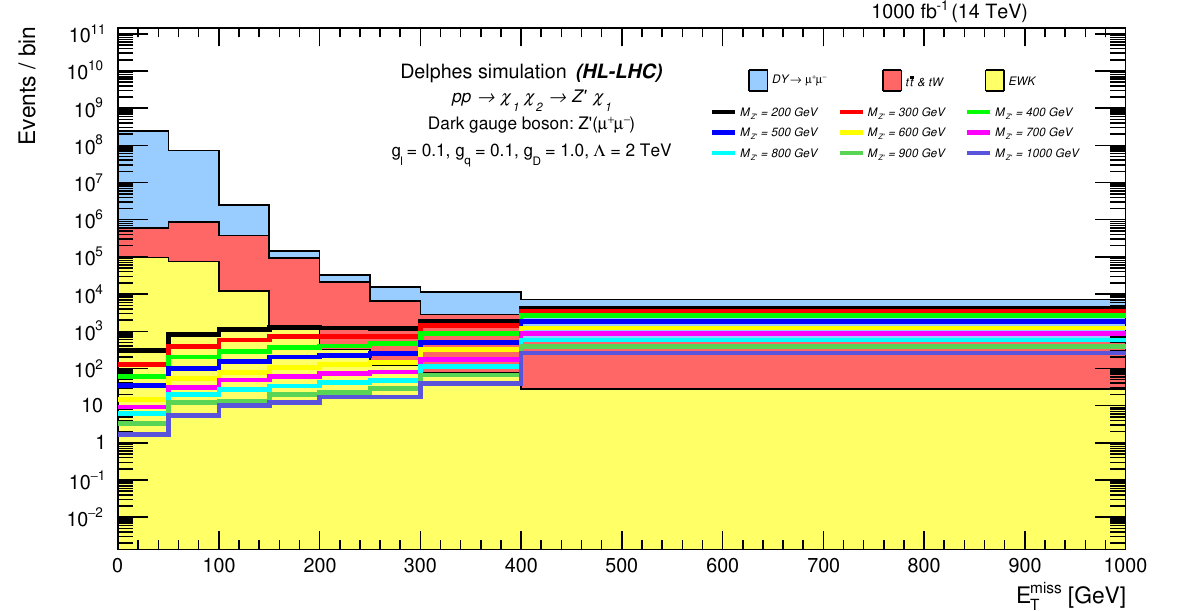}
  \label{met}
}
\caption{The histograms of (a) $M_{\mu\mu}$ and (b) the $E_{T}^{miss}$ distributions for the expected SM background and the EFT scenario signal with different $M_{Z^{\prime}}$ values in the HDS at $\Lambda = 2$ TeV. The distributions are shown after applying the preliminary selection criteria listed in table \ref{table:tab3}.}
\label{figure:fig3}
\end{figure*}
In this analysis, the event selection is designed to reconstruct two high-$p_{T}$ muons of opposite charges associated with $E_{T}^{miss}$, referring to DM candidates, in the final state. This selection is presented as applied cuts to different kinematics requiring muons to pass the preliminary selection shown in table \ref{table:tab3}.

Thus, each muon should be with $p^{\mu}_{T} > 30$ GeV and $|\eta^{\mu}| < 2.4$, in addition, this muon must be isolated and pass the following criteria "IsolationVarRhoCorr" referring to the isolation cut in Delphes to eliminate muons produced inside jets. For this cut, corrected for the pileup effect, the scalar summation of the $p_{T}$ of all the tracks of muons should not go beyond 10$\%$ of the muon $p^{\mu}_{T}$. 
This must be in the limits of a $\Delta$R = 0.5 cone surrounding the muon candidate, except the candidate itself. $M_{\mu\mu}$ is greater than 60 GeV as we search for a high mass regime resonance.
\begin{table*}
    \centering
\begin{tabular}{| c | c | c |} 
 \hline
 Step & Criteria & Requirements \\ [1.0ex] 
 \hline\hline
  & $p_{T}^{\mu}$ (GeV) & $> 30$ \\ 
 Preliminary selection & $|\eta^{\mu}|$ (rad) & $< 2.4$ \\ 
  & IsolationVarRhoCorr & $< 0.1$ \\
  & $M_{\mu^{+}\mu^{-}}$ (GeV) & $> 60$ \\
 \hline
  & Mass window (GeV) & $0.9 \times M_{Z^{\prime}}$ $< M_{\mu^{+}\mu^{-}} <$ $M_{Z^{\prime}} + 25$\\
  Tight selection & $|P_{T}^{\mu^{+}\mu^{-}} - E_{T}^{miss}|/P_{T}^{\mu^{+}\mu^{-}}$ & $<0.4$ \\
  & $\Delta R^{\mu^{+}\mu^{-}}$ & $<3.2$ \\
  & $\Delta\phi_{\mu^{+}\mu^{-},E_{T}^{miss}}$ & $>2.6$\\[1ex] 
 \hline
\end{tabular}
\caption{Summary of the preliminary and tight cuts of the final event selection used in this cut-based analysis.}
\label{table:tab3}
\end{table*}

Figure \ref{mass} displays the histograms of $M_{\mu\mu}$ distribution for the signal and the stacked background histograms. The blue histogram, the dominant background, refers to the DY background, while the yellow histogram is for the EW background of vector-boson pairs. The backgrounds of the top-quark pair, and the single-top are denoted by the red histogram. The different colored lines overlaid on the stacked histograms of the background represent the EFT signal, generated for various values of $M_{Z^{\prime}}$, in the HDS masses assumptions at $\Lambda$ = 2 TeV. The corresponding $E_{T}^{miss}$ distribution is illustrated in figure \ref{met}.


In figures \ref{mass} and \ref{met}, as the background overwhelms the signal, we need a tighter set of discrimination cuts to distinguish between the signals and SM backgrounds. We apply four cut parameters: 
The first is the requirement of $M_{\mu\mu}$ to be restricted to a small range of the $M_{Z^{\prime}}$, where $0.9 \times M_{Z^{\prime}}$ $< M_{\mu^{+}\mu^{-}} <$ $M_{Z^{\prime}} + 25$ as recommended by \cite{R1}. 
The second is the selection of the relative difference of the dimuon transverse momentum ($P_{T}^{\mu^{+}\mu^{-}}$) and the $E_{T}^{miss}$ to be less than 0.4 (i.e. $\mid P_{T}^{\mu^{+}\mu^{-}}$ - $E_{T}^{miss} \mid/P_{T}^{\mu^{+}\mu^{-}} < 0.4$). 
The third is the cone radius $\Delta R^{\mu^{+}\mu^{-}}$ to be less than 3.2 (i.e. $\Delta R^{\mu^{+}\mu^{-}} <$ 3.2). 
The fourth is the azimuthal angle between the dimuon and the $E_{T}^{miss}$ directions, defined as $\Delta$$\phi_{\mu^{+}\mu^{-},E_{T}^{miss}} = \mid \Phi^{\mu^{+}\mu^{-}} - \Phi^{miss} \mid$, to be selected as $\Delta$$\Phi_{\mu^{+}\mu^{-},E_{T}^{miss}} > 2.6$. 
Table \ref{table:tab3} summarizes these selection criteria.

\begin{figure}
\centering
\subfigure[$|P_{T}^{\mu^{+}\mu^{-}} - E_{T}^{miss}|/P_{T}^{\mu^{+}\mu^{-}}$]{ 
\includegraphics[width=85mm]{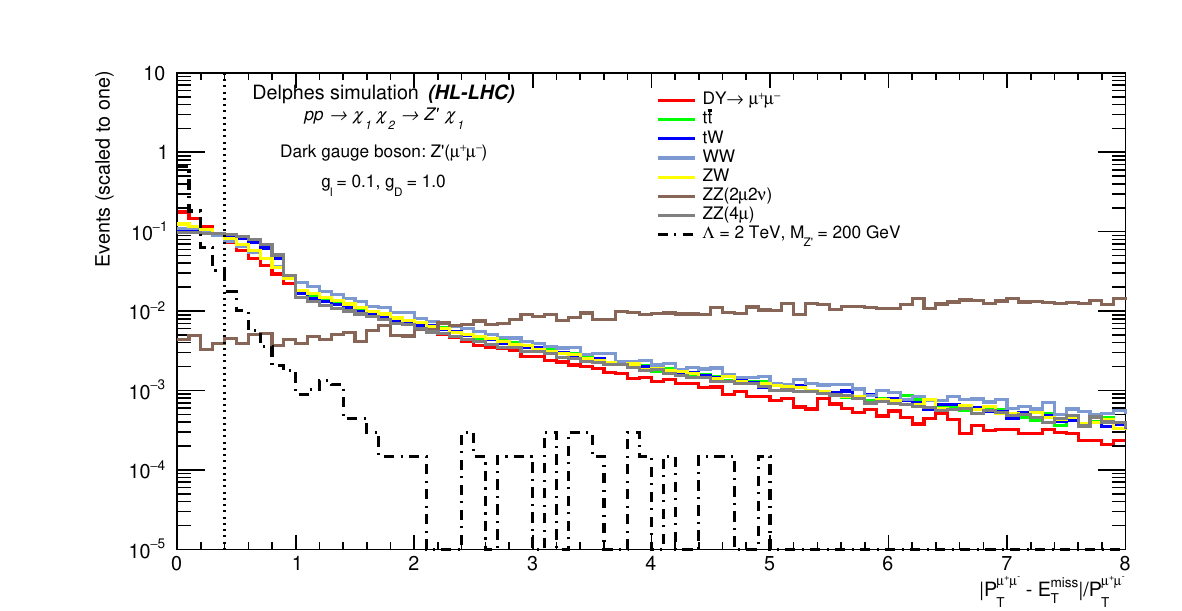}
\label{cut1}
}
\hspace{0mm}
\subfigure[$\Delta R^{\mu^{+}\mu^{-}}$]{
  \includegraphics[width=85mm]{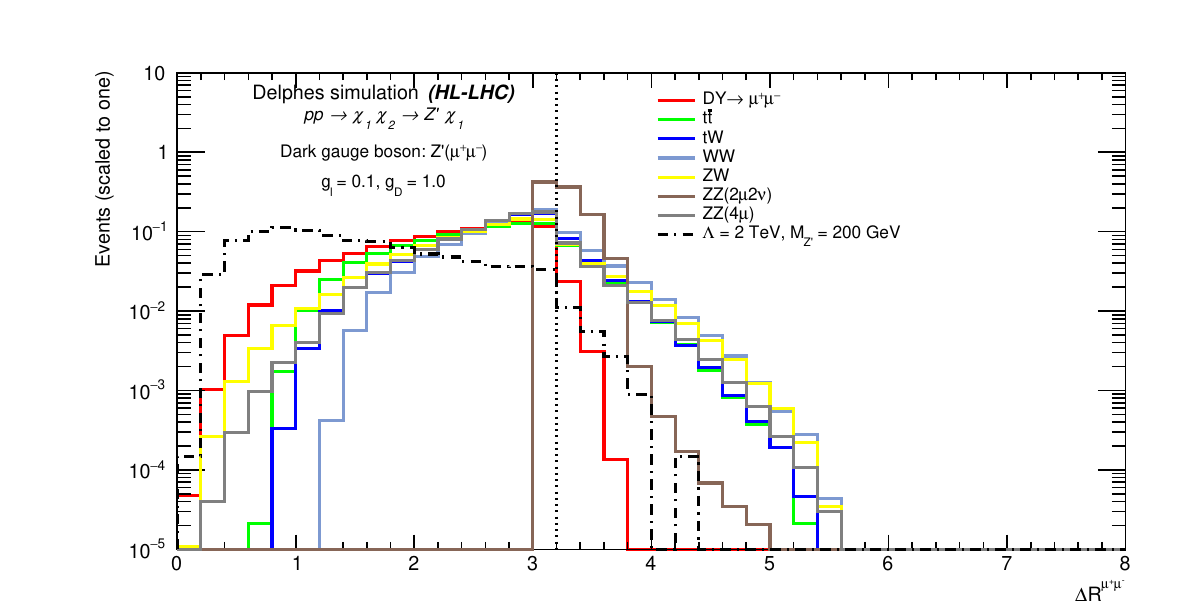}
  \label{cut2}
}
\hspace{0mm}
\subfigure[$\Delta\phi_{\mu^{+}\mu^{-},E_{T}^{miss}}$]{
  \includegraphics[width=85mm]{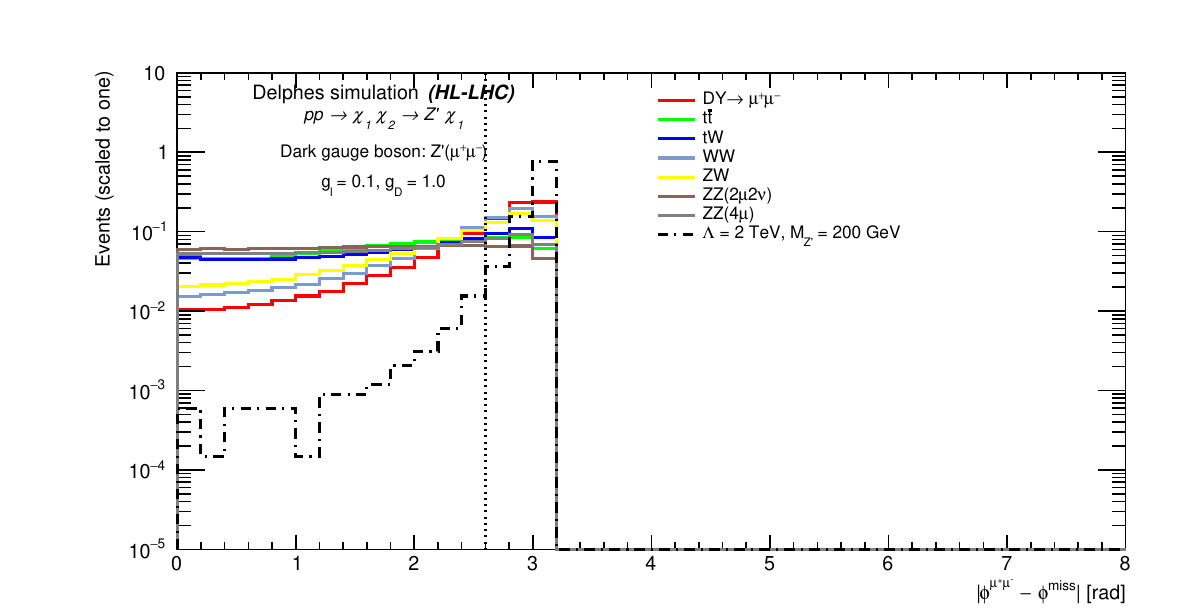}
  \label{cut3}
}
\caption{ Normalized histograms for EFT scenario signal ($\Lambda = 2$ TeV, $M_{Z^{\prime}} = 200$ GeV) and SM backgrounds. The vertical dashed lines indicate the chosen cut values for each variable.}
\label{figure:cuts}
\end{figure}

Figure \ref{figure:cuts} shows the distributions, scaled to one, of $|P_{T}^{\mu^{+}\mu^{-}}$ - $E_{T}^{miss}|/P_{T}^{\mu^{+}\mu^{-}}$ \ref{cut1}, $\Delta R^{\mu^{+}\mu^{-}}$ \ref{cut2}, and $\Delta\phi_{\mu^{+}\mu^{-},E_{T}^{miss}}$ \ref{cut3} with their corresponding cutting values for dimuons events applied to the SM background and the generated signal of EFT scenario in HDSr for $M_{Z^\prime}$ = 200 GeV and $\Lambda$ = 2 TeV. The application of the mass window cut ($0.9 \times M_{Z^{\prime}}$ $< M_{\mu^{+}\mu^{-}} <$ $M_{Z^{\prime}} + 25$) fully suppresses the $ZZ\rightarrow{\mu^{+}\mu^{-} + 2\nu}$ background.

The next section presents the results of applying these cuts to the $E_{T}^{miss}$ showing how strongly these cuts reduced the SM background for the sake of discrimination between the signal and the SM background.
\section{Results}
\label{section:Results}
\begin{figure}[]
\centering
\resizebox*{85mm}{!}{\includegraphics{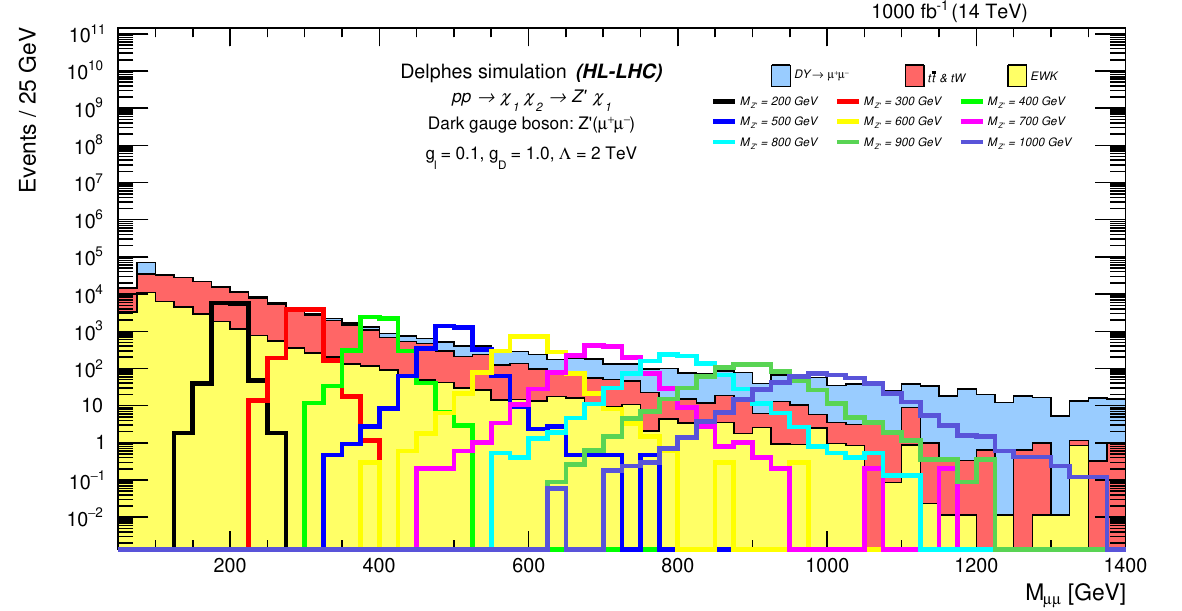}}
\caption{ $M_{\mu\mu}$ distributions for the expected SM backgrounds and EFT scenario signal with $M_{Z^{\prime}} = 200$ GeV in HDS at $\Lambda = 2$ TeV. The distributions are shown after applying all selection cuts listed in table \ref{table:tab3}, excluding the mass window cut.} 
\label{figure:fig5}
\end{figure}

The shape-based analysis is chosen based on the $E_{T}^{miss}$ distribution because of the good discrimination it gives between the model's signal and the SM background combination. Figure \ref{figure:fig5} displays $M_{\mu\mu}$ distribution, related to the production of Z$^{\prime}$, after applying the preliminary and tight cuts, listed in table \ref{table:tab3}, except the mass window cut.
As noticed, there is a significant decrease in SM backgrounds while preserving the signal strength, as demonstrated by the comparison of figures \ref{mass} and \ref{figure:fig5} for the EFT scenario.
\begin{figure}[]
\centering
\resizebox*{85mm}{!}{\includegraphics{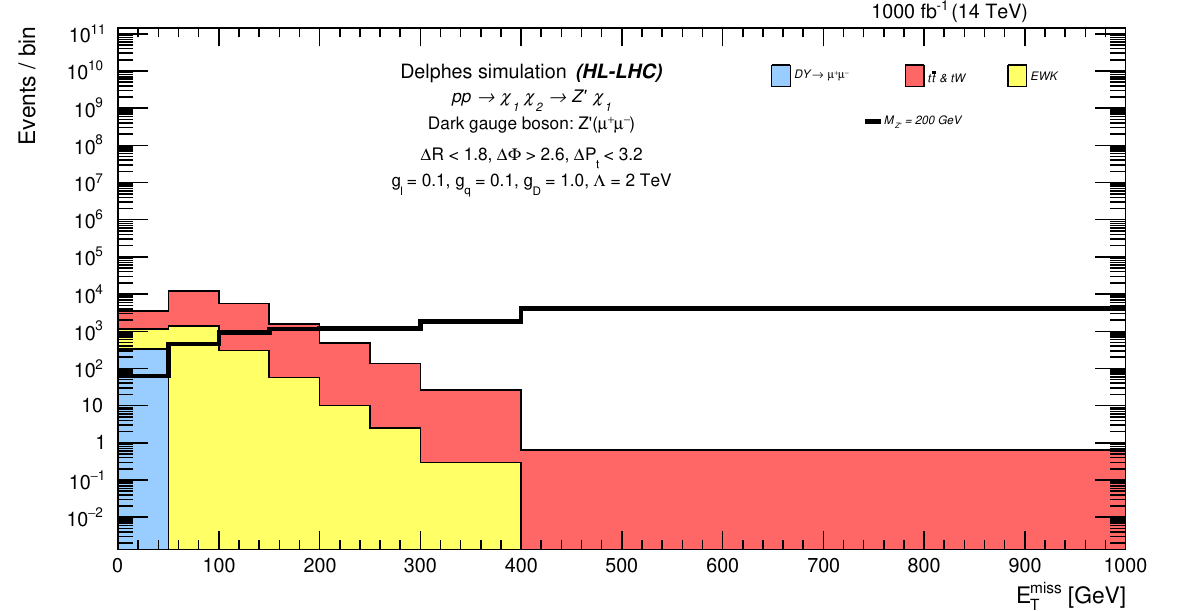}}
\caption{$E_{T}^{\text{miss}}$ distribution for the expected SM backgrounds and EFT scenario signal with $M_{Z^{\prime}} = 200$ GeV in HDS at $\Lambda = 2$ TeV. The distribution is shown after applying all selection cuts listed in table \ref{table:tab3}.} 
\label{figure:fig6}
\end{figure}
Figure \ref{figure:fig6} illustrates $E_{T}^{miss}$ distribution after applying all the cuts presented in table \ref{table:tab3}.  

We utilized the profile likelihood method to statistically analyze our findings and to conduct a statistical test based on $E_{T}^{miss}$ distributions. Employing the modified frequentist construction confidence levels (CL) \cite{R60, R61}, which is grounded on the asymptotic approximation \cite{R62}, we determined exclusion limits $\sigma\times Br(Z'\rightarrow \mu\mu)$ of the signal at a 95\% CL.
 
\begin{figure*}
\centering
    \resizebox*{11.0cm}{!}{\includegraphics[width=75mm]{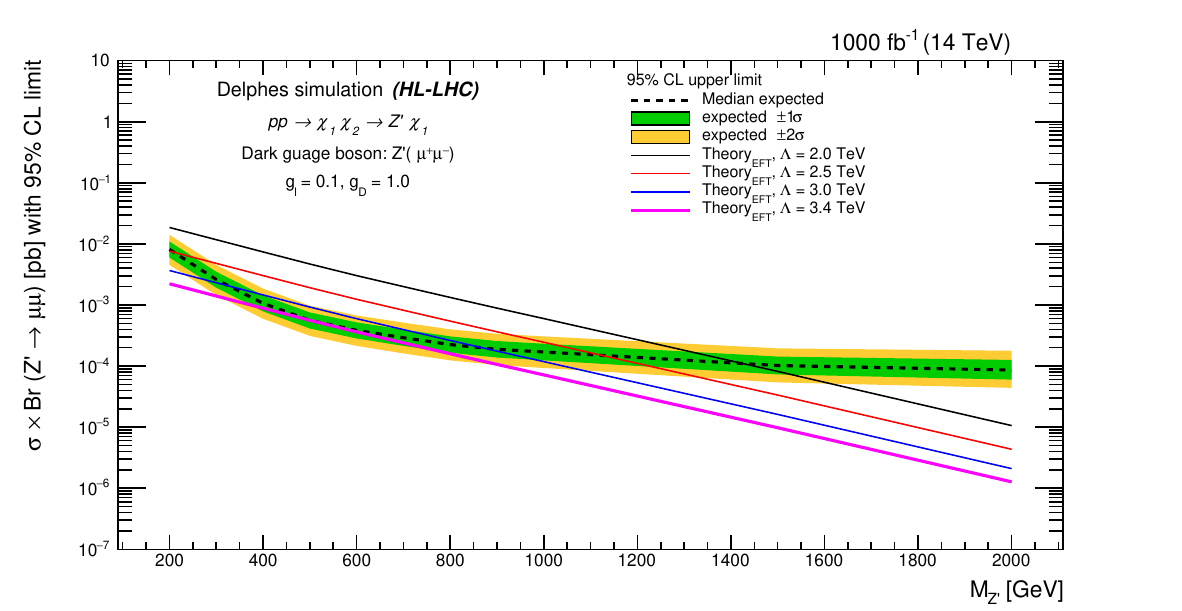}}
    \hspace{0mm}
    \caption{95\% CL upper limits on $\sigma \times \text{Br}(Z' \rightarrow \mu\mu)$ (expected) as a function of the $M_{Z^{\prime}}$ in EFT scenario, assuming muonic decay of the $Z^{\prime}$. The solid lines represent the EFT scenario for different values of $\Lambda$: $\Lambda = 2$ TeV (black), 2.5 TeV (red), 3 TeV (blue), and 3.4 TeV (pink).}
   \label{figure:limits}
\end{figure*}

Figure \ref{figure:limits} illustrates the 95$\%$ CL upper limit on $\sigma\times Br(Z'\rightarrow \mu\mu)$ plotted against $M_{Z^{\prime}}$ for EFT scenario based on the mono-Z$^{\prime}$ portal. This specifically pertains to the muonic decay of Z$^{\prime}$ and for ${g}_{D} = 1.0$, ${g}_{l} = 0.1$ in the HDS, as detailed in table \ref{table:tab1}, for various $\Lambda$ values.

The solid curves in this plot represent the theoretical $\sigma\times Br(Z'\rightarrow \mu\mu)$ corresponding to the EFT scenario for specific values of $\Lambda$. 
Based on figure \ref{figure:limits}, the production of Z$^{\prime}$ is ruled out in the mass range of 200 to 1420 GeV for $\Lambda =$ 2 TeV, and ruling out $M_{Z'}$ equal to 512 GeV for $\Lambda =$ 3.4 TeV as indicated from the expected median.
Above $\Lambda =$ 3.4 TeV the HL-LHC is not sensitive to the EFT scenario. 
\begin{figure*}
\centering
  \resizebox*{11.0cm}{!}{\includegraphics{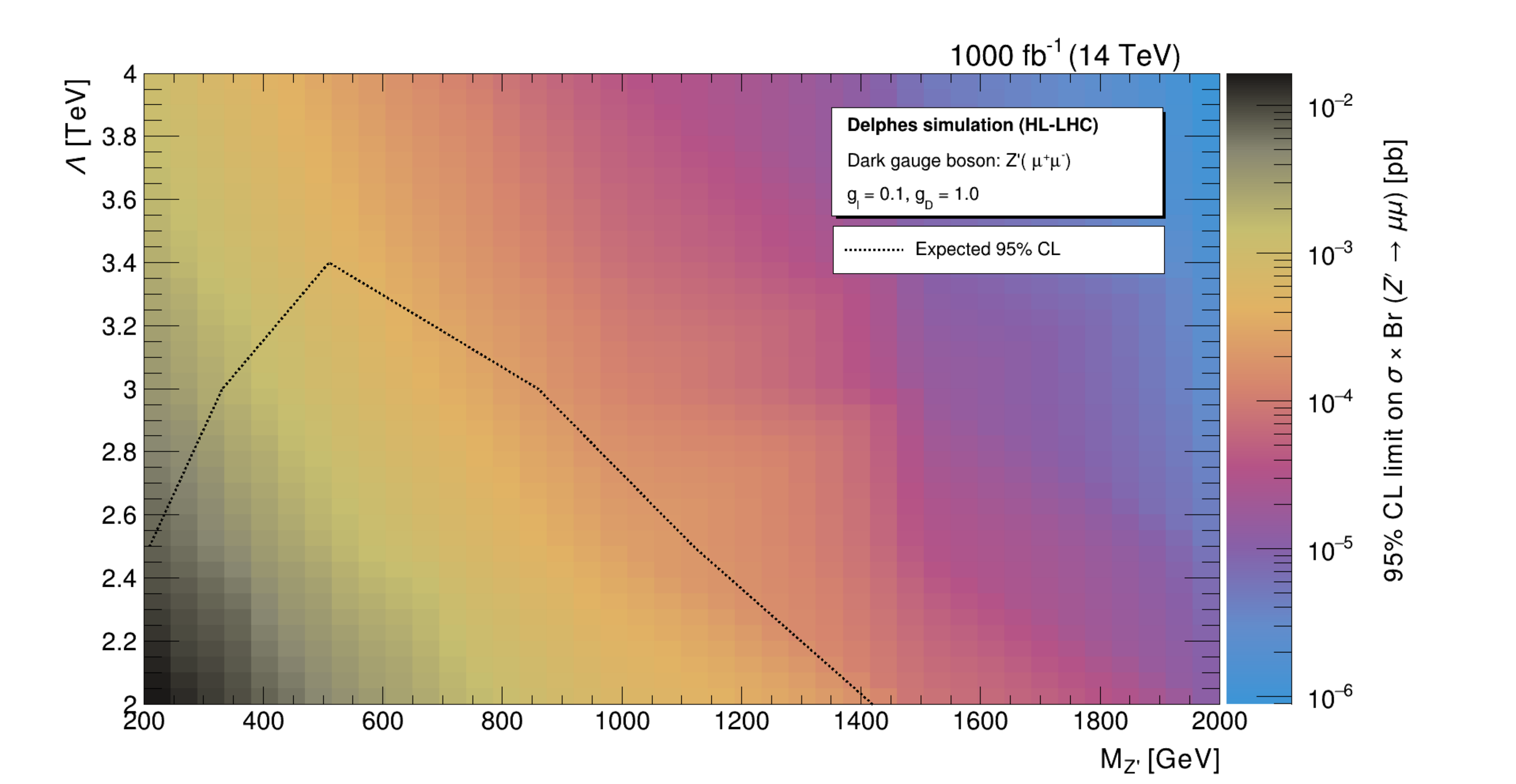}}
   \caption{95\% CL upper limits on $\sigma \times \text{Br}(Z' \rightarrow \mu\mu)$ from the inclusive search, evaluated for different pairs of EFT scenario parameters, $\Lambda$ and $M_{Z^{\prime}}$. The colored region represents the upper limit of 95\% CL, while the dashed black line indicates the expected exclusions for the nominal $Z^{\prime}$ cross-sections.}
  \label{figure:2Dlimit}
\end{figure*} 

Figure \ref{figure:2Dlimit} presents the anticipated exclusion limits in a 95\% CL resulting from EFT scenario search with $\mathcal{L}$ = 1000 fb$^{-1}$ and $\sqrt{s}$ = 14 TeV proton-proton collision simulated data. These limits are presented as a function of $M_{Z^{\prime}}$ and $\Lambda$. The colored region represents the upper limit of 95\% CL, while a dashed black line denotes the median of the expected limits. The area under the expected limit curve is excluded.

\section{Summary}
\label{section:Conclusion}
This study proposed a search for a Z$^{\prime}$ boson decaying into dimuon in association with neutral particles (DM particles $\chi_{1}$ and $\chi_{2}$) in framework of the EFT scenario.

MC simulations of proton-proton collisions at $\sqrt{s}$ = 14 TeV for $\mathcal{L}$ = 1000 fb$^{-1}$, corresponding to the HL-LHC upgrade, were used. The analysis presented results of the muonic decay of Z$^{\prime}$ for HDS ($M_{\chi_{1}} = M_{Z^{\prime}}/2$, and $M_{\chi_{2}} = 2M_{Z^{\prime}}$), given that the coupling constants are fixed and chosen to be $g_{l} = 0.1$ and $g_{D} = 1.0$. 
95\% CL upper limit on the free parameters of the model is presented for $\Lambda$ and $M_{Z^{\prime}}$.

If the signal is not detected at the HL-LHC with $\sqrt{s}$ = 14 TeV, we establish a 95\% CL upper limits on $M_{Z^{\prime}}$ and $\Lambda$ for the charged muonic decay channel of Z$^{\prime}$. Specifically, limits have been determined for EFT scenario with $g_{l} = 0.1$ and $g_{D} = 1.0$, ruling out $M_{\mu\mu}$ range from 200 to 1420 GeV for $\Lambda \in [2, 3.4]$ TeV, while notably excluding $\Lambda = 3.4$ TeV at $M_{Z^{\prime}} = 512$ GeV. Furthermore, for exceedingly high values of $\lambda$ (i.e., $> 3.4$ TeV), the HL-LHC will lack sensitivity to the EFT scenario.
\section{Acknowledgments}
\label{section:Acknowledgments} 
The author of this paper appreciates the help of Tongyan Lin, an author in \cite{R1}, for sending us the Universal FeynRules Output (UFO) files for the model used in the generation of the events. We also thank Mahmoud Hashim for his help with the IT issues faced throughout this work.


\end{document}